\documentclass{bmvc2k}

\usepackage{epsfig}
\usepackage{amsmath}
\usepackage{amssymb}
\usepackage{multirow}
\usepackage{adjustbox}
\usepackage{booktabs}
\usepackage{enumitem}
\usepackage{graphicx}
\usepackage{caption}
\usepackage{wrapfig}
\usepackage{subcaption} 
\title{Adversarial Attacks on Audio Deepfake Detection: A Benchmark and Comparative Study}
\addauthor{Kutub Uddin}{kutub@umich.edu}{1}
\addauthor{Muhammad Umar Farooq}{mufarooq@umich.edu}{1}
\addauthor{Awais Khan}{mawais@umich.edu}{1}
\addauthor{Khalid Mahmood Malik}{drmalik@umich.edu}{1}

\addinstitution{
College of Innovation \& Technology\\
University of Michigan-Flint\\
MI, 48502, USA\\
}

\runninghead{}{BMVC}

\begin{document}

\maketitle
\vspace{-15pt}
\begin{abstract}\vspace{-3pt}
The widespread use of generative AI has shown remarkable success in producing highly realistic deepfakes, posing a serious threat to various voice biometric applications, including speaker verification, voice biometrics, audio conferencing, and criminal investigations. To counteract this, several state-of-the-art (SoTA) audio deepfake detection (ADD) methods have been proposed to identify generative AI signatures to distinguish between real and deepfake audio. However, the effectiveness of these methods is severely undermined by anti-forensic (AF) attacks that conceal generative signatures. These AF attacks span a wide range of techniques, including statistical modifications (e.g., pitch shifting, filtering, noise addition, and quantization) and optimization-based attacks (e.g., FGSM, PGD, C \& W, and DeepFool). In this paper, we investigate the SoTA ADD methods and provide a comparative analysis to highlight their effectiveness in exposing deepfake signatures, as well as their vulnerabilities under adversarial conditions. We conducted an extensive evaluation of ADD methods on five deepfake benchmark datasets using two categories: raw and spectrogram-based approaches. This comparative analysis enables a deeper understanding of the strengths and limitations of SoTA ADD methods against diverse AF attacks. It does not only highlight vulnerabilities of ADD methods, but also informs the design of more robust and generalized detectors for real-world voice biometrics. It will further guide future research in developing adaptive defense strategies that can effectively counter evolving AF techniques.
\end{abstract}\vspace{-15pt}

%-------------------------------------------------------------------------
\section{Introduction}\vspace{-10pt}
\label{sec:intro}
The rapid advancement of generative AI has significantly increased the accessibility of deepfakes~\cite{todisco2019asvspoof} to mimic human voices. Although deepfakes improve user convenience in voice-biometrics, including smart assistants, such as Amazon Alexa, Google Home, and Apple Siri, they also introduce serious risks~\cite{netinant2024development}. In particular, attackers can use deepfakes to impersonate users and bypass voice biometrics~\cite{khan2024parallel}. Consequently, systems ranging from automatic speaker verification to ADDs are facing security challenges. Recent analyses from industry and academia reports reveal a sharp increase in deepfake-related incidents~\cite{pindrop2025report}. In 2024 alone, an estimated \$12.5 billion was lost to fraud related to AI-driven attacks, with approximately \$2.6 million of that loss specifically attributed to audio deepfakes~\cite{pindrop2025visr}.\\
To address these risks, the research community has proposed several ADDs~\cite{10.1007}. These ADDs are based on raw audio or acoustic features (e.g., MFCC, LFCC, Spectrogram) with traditional classifiers~\cite{10374968,10.1145,10447500}, end-to-end deep learning~\cite{jung2022pushing,tak2022rawboost,tak2021end}, and more recent fine-tuned foundation models~\cite{10.1007}. In particular, ADDs aim to capture subtle artifacts from raw or spectrogram signals during generation. However, as ADDs become more sophisticated, they have become increasingly vulnerable to AFs. AFs span a diverse techniques, from statistical (e.g., noise injection~\cite{li2025measuring}), to recent optimization (e.g., FGSM~\cite{goodfellow2014explaining}, PGD~\cite{madry2017towards}, C {\&} W~\cite{carlini2017towards}, and DeepFool~\cite{moosavi2016deepfool}) that can significantly degrade the performance of ADDs.\\
Although robustness has been extensively explored in the image domain~\cite {alghamdi2024enhancing, uddin2019anti, park2024comprehensive, uddin2024counter, uddin2023robust, uddin28enhanced}, its application to ADD remains relatively unexplored. Some recent studies~\cite{farooq2025transferable,kawa2023,rabhi2024audio,wu2024clad,uddin2025shieldsecurehighlyenhanced} have investigated the vulnerability of ADDs and demonstrated the performance degradation by AFs. For example, Wu et al.~\cite{wu2024clad} presented contrastive learning-based ADD to improve robustness against AFs, such as volume control, fading, and noise injection. Although effective against these perturbations, the study~\cite{wu2024clad} does not address complex AFs such as optimization~\cite{goodfellow2014explaining, uddin2023deep}. Similarly, studies such as~\cite{farooq2025transferable,kawa2023} are limited in both architectural scope and dataset diversity. Furthermore, a recent survey~\cite{rabhi2024audio} reviews AFs on audiovisual deepfake, including defense methods such as fusion and decoy techniques, but lacks comparative evaluations of vulnerabilities against AF attacks. \\
Despite progress~\cite{rabhi2024audio,farooq2025transferable}, a critical gap remains in unified evaluation of ADDs under diverse AFs across various designs and datasets. Most ADDs~\cite{jung2022pushing,khan2024parallel,tak2022rawboost,tak2021end} focus on limited perturbations (e.g., noise) or test on selected datasets without cross-corpus evaluation. Thus, current evaluations miss AF scenarios where attackers exploit ADDs to evade them.\\
To our knowledge, this is the first large-scale comparative study of AFs on both raw and spectrogram-based ADDs across diverse datasets and AF types. Due to the lack of standardized evaluation, it remains unclear which methods are more robust. We address this gap through a unified, cross-architecture, and cross-dataset analysis to offer a comprehensive study across input formats, ADDs, and AFs. Our key contributions are:
\begin{itemize}[noitemsep,topsep=0pt]
    \item We present the first unified, large-scale, and apple-to-apple comparative evaluation of SoTA ADDs under AF attacks, focused on voice biometric applications.
    \item We benchmark twelve ADDs across five datasets under two AF categories: statistical (e.g., noise) and optimization-based (e.g., FGSM, PGD, C \& W, DeepFool) attacks.
    \item We provide an in-depth analysis of current methods regarding key limitations and suggesting insights for designing more robust and generalizable ADD methods.
\end{itemize}\vspace{-15pt}

\section{Benchmarks Selection} \vspace{-10pt}
\label{sec:benchmarks}
This section describes the datasets, ADDs, and AF techniques for the comparative study.
\\\vspace{-28pt}
\subsection{Datasets} \vspace{-8pt}
Based on the recent benchmark study, we selected five widely used datasets to assess the performance and vulnerabilities of ADD systems. The selected datasets are as follows: \\
\textbf{ASVSpoof2019 ($D1$)}~\cite{todisco2019asvspoof}: The ASVSpoof2019 dataset~\cite{todisco2019asvspoof} includes logical and physical access tracks. We use a logical access subset that contains over 121,000 utterances generated using 17 TTS and VC systems. \\
\textbf{ASVSpoof2021 ($D2$)}~\cite{liu2023asvspoof}: The ASVSpoof2021 dataset~\cite{liu2023asvspoof} builds on previous editions~\cite{todisco2019asvspoof} with a wider set of spoofing attacks on logical and physical access tracks and comprises more than 181,000 utterances.\\
\textbf{ASVSpoof2024 ($D3$)}~\cite{wang2024asvspoof}: The ASVSpoof2024 dataset~\cite{wang2024asvspoof} is the most recent in the ASVSpoof series, introducing multilingual, cross-device, and adversarial spoofing conditions. It comprises 182,000 training, 140,000 development, and 680,000 test samples. \\
\textbf{CodecFake ($D4$)}~\cite{xie2025codecfake}: The CodecFake dataset~\cite{xie2025codecfake} includes deepfakes generated by Audio Language Models (ALMs), which combine language modeling and codec compression to synthesize high-fidelity speech. \\
\textbf{WaveFake ($D5$)}~\cite{frank2021wavefake}: The WaveFake dataset~\cite{frank2021wavefake} is a large-scale dataset that includes over 117,000 clips synthesized using state-of-the-art vocoders and GANs, offering diverse generation techniques for ADD benchmarking.\\ \vspace{-25pt}

\begin{figure*}[!b]
    \centering
    \vspace{-15pt}
    \includegraphics[width=0.9\linewidth]{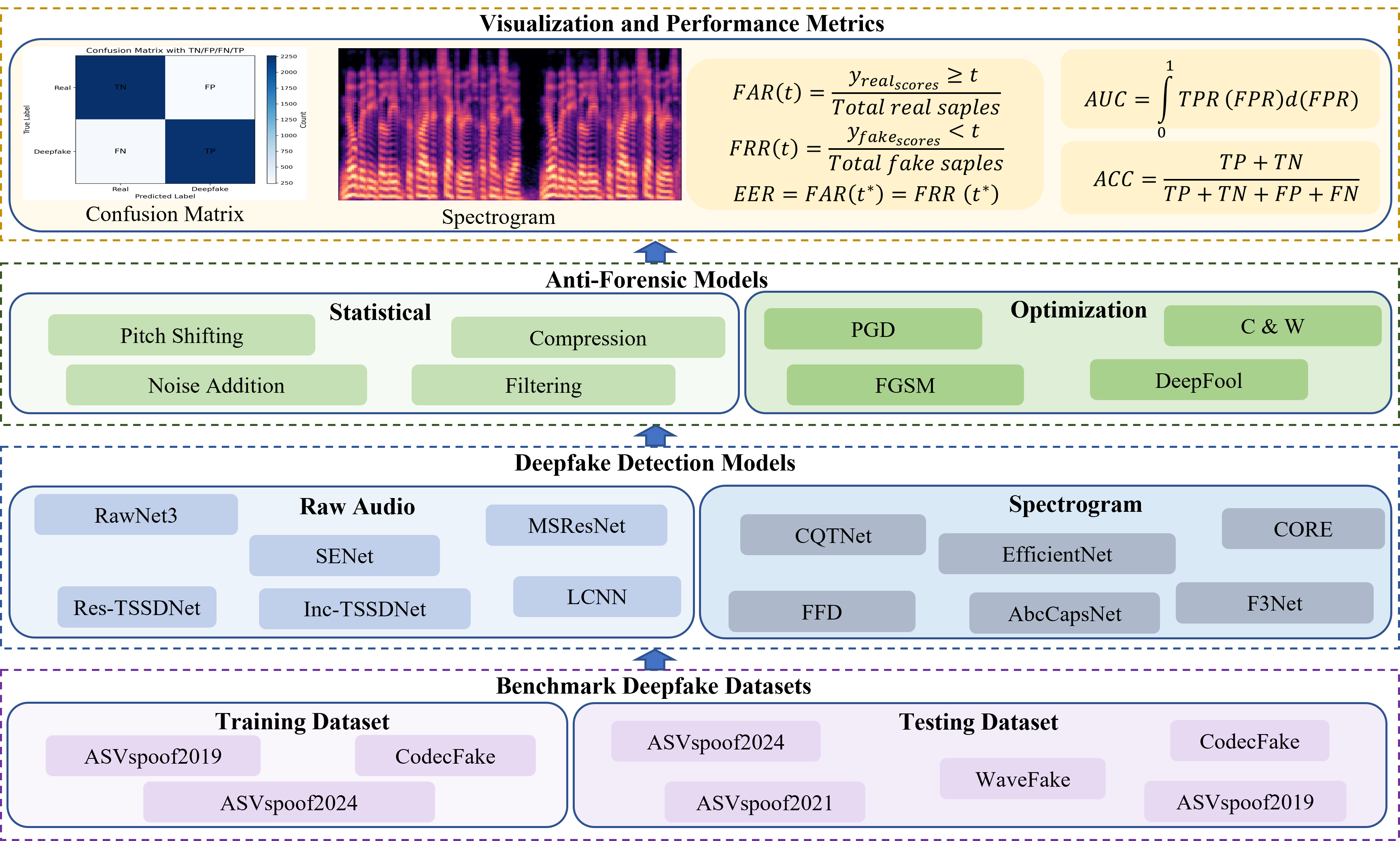} 
    \vspace{7pt}
    \caption{Architecture of the proposed comparative study. It starts with dataset selection, followed by two ADD groups, such as raw and spectrogram-based, to reveal synthetic traits. Two AF attack types, such as statistical and optimization-based, are applied to deceive the ADDs. Visualizations offer perceptibility and qualitative analysis across AF attack types.}   
    \label{fi:arch}
\end{figure*}

\subsection{Deepfake Detection Methods} \vspace{-8pt}
\label{sec:dfd}
We consider two principal categories of SoTA ADD models: (a) raw audio and (b) spectrogram-based methods. These methods were chosen based on their open-source availability, proven effectiveness, and representation of diverse architectural designs. \vspace{-13pt}

\subsubsection{Raw Audio-based Methods} \vspace{-8pt}
Raw ADDs operate directly on time-domain signals, learning temporal and low-level patterns to distinguish real and deepfakes. In our benchmark, we include six SoTA raw ADDs, including RawNet3 ($R1$)~\cite{jung2022pushing}, MS-ResNet ($R2$)~\cite{wang2018csi}, SeNet ($R3$)~\cite{wu2020defense}, LCNN ($R4$)~\cite{kawa2022defense}, Res-TSSDNet ($R5$)~\cite{hua2021towards}, and Inc-TSSDNet ($R6$)~\cite{hua2021towards}. In particular, RawNet3 ($R1$)~\cite{jung2022pushing} integrates the Res2Net backbone and multi-layer feature aggregation, which shows strong performance even with limited labeled data.
MS-ResNet ($R2$)~\cite{wang2018csi} and SENet ($R3$)~\cite{wu2020defense} are both ResNet-based variants; the MS-ResNet ($R2$)~\cite{wang2018csi} incorporates multi-scale residual blocks to extract spatial patterns, while the SENet ($R3$)~\cite{wu2020defense} enhances robustness via adversarial training and spatial smoothing. LCNN ($R4$)~\cite{kawa2022defense} introduces lightweight convolutional layers to learn discriminative features from raw waveforms. Res-TSSDNet ($R5$)~\cite{hua2021towards} and Inc-TSSDNet ($R6$)~\cite{hua2021towards} are lightweight, end-to-end models optimized for ADD. Res-TSSDNet ($R5$)~\cite{hua2021towards} uses a ResNet-style architecture with stacked residual blocks, whereas Inc-TSSDNet ($R6$)~\cite{hua2021towards} employs an Inception with parallel dilated convolutions to expand receptive fields. Both demonstrate strong generalization across datasets.
\vspace{-13pt}
\subsubsection{Spectrogram-based Methods} \vspace{-8pt}
Spectrogram-based ADDs first transform audio into time-frequency (e.g., Mel or CQT spectrograms), then capture visual patterns to detect deepfake artifacts. Our benchmark includes six SoTA spectrogram-based ADDs, including ABCCapsNet ($S1$)~\cite{wani2024abc}, EfficientNet ($S2$)~\cite{tan2019efficientnet}, FFD ($S3$)~\cite{dang2020detection}, CORE ($S4$)~\cite{ni2022core}, F3Net ($S5$)~\cite{qian2020thinking}, and CQTNet ($S6$)~\cite{ziabary2021countermeasure}. \\
ABCCapsNet ($S1$)~\cite{wani2024abc} combines attention-based bottleneck features with capsule networks to capture spatial relationships, and offers robustness against temporal shifts and AF perturbations.
EfficientNet ($S2$)~\cite{tan2019efficientnet} applies compound scaling to balance detection accuracy and model efficiency to enable effective capture of localized spectral distortions. FFD ($S3$)~\cite{dang2020detection} employs attention mechanisms to highlight manipulated regions within spectrogram images to detect deepfake traces. CORE~\cite{ni2022core} and FFD~\cite{dang2020detection} generalize well to spectrogram-based ADD due to their architecture's focus on structural inconsistencies. CORE ($S4$)~\cite{ni2022core} introduces a consistency loss across augmented views to enforce invariant learning, thereby emphasizing intrinsic forgery cues over superficial artifacts. F3Net ($S5$)~\cite{qian2020thinking} integrates frequency decomposition with local frequency statistics to enhance detection performance. CQTNet ($S6$)~\cite{ziabary2021countermeasure} employs Constant-Q Transform spectrograms and self-attended ResNet blocks with one-class learning to improve generalization to unseen attacks. \vspace{-13pt}

\subsection{Anti-Forensic Methods} \vspace{-8pt}
\label{sec:af_methods}
We categorize the selected AF attacks into two primary classes: (a) statistical-based and (b) optimization-based methods. This taxonomy encompasses a representative range of AF strategies frequently observed in real-world deepfake generation and evasion scenarios. \vspace{-12pt}

\subsubsection{Statistical AF Attacks} \vspace{-8pt}
\label{sec:st_af}
Statistical AF attacks, such as pitch shifting~\cite{chirkovskiy2025pitch}, filtering~\cite{wu2020defense}, noise addition~\cite{li2025measuring}, and compression~\cite{li2025measuring}, aim to alter the statistical properties, such as amplitude distribution, spectral content, or temporal patterns, without noticeably changing their perceptual quality. The primary goal is to evade ADD models that rely on these cues to identify synthetic nature.\\
\textbf{Pitch Shifting~\cite{chirkovskiy2025pitch}} is a statistical AF method that alters amplitude distribution, spectral content, or temporal patterns without changing the audio's duration. It is often overlooked in deepfake detection, posing a potential vulnerability. We apply a phase vocoder for time-stretching, followed by resampling to shift the pitch as follows:
\vspace{-8pt}
\begin{align}
A' &= \mathcal{F}^{-1} \left\{ \mathcal{F}\big(A\big) \cdot H(f, n) \right\} 
\end{align}
Here, ${F}$ and ${F}^{-1}$ denote the Fourier and inverse transforms of frequency $f$ by n semitones. We apply pitch shifts of ±1, ±5, and ±12 semitones to simulate lowering and raising effects.\\
\textbf{Filtering~\cite{wu2020defense}} is a technique used to remove or isolate specific frequency components of an audio signal. In ADD, median filtering can be used to simulate deepfake artifacts to create adversarial perturbations, defined as follows:
\vspace{-7pt}
\begin{equation}
A' = \frac{A\left( \frac{N}{2} \right) + A\left( \frac{N}{2} + 1 \right)}{2}
\end{equation}
We apply kernel sizes $N \in {3, 5, 7, 9}$ to control the strength of the AF perturbations.\\
\textbf{Noise Addition~\cite{li2025measuring}} is a statistical approach that degrades the statistical patterns and spectral features of audio to obscure deepfake artifacts. We applied Gaussian noise as a statistical AF attack, defined as follows:
\vspace{-6pt}
\begin{equation}
A' = A + \delta; \quad \delta \sim \mathcal{N}(0, \sigma^2) 
\end{equation}
We set the standard deviation, $\sigma$, as 0.001, 0.01, 0.02, 0.03, 0.04, and 0.05.\\
\textbf{Quantization~\cite{grama2015quantization}} is a statistical transformation that reduces the precision of an audio by mapping its continuous amplitudes to a limited number of discrete levels. This process alters the fine-grained amplitude distribution and introduces quantization noise, thereby distorting the subtle statistical patterns that deepfake detectors often rely on, such as high-frequency details, spectral smoothness, and dynamic range. This is computed as follows:
\vspace{-8pt}
\begin{equation}
A' = \frac{
    \mathrm{round}\left( \left(A + 1 \right) \times \left(\frac{L}{2} - 1\right) \right)
}{
    \left(\frac{L}{2} - 1\right)
} - 1
\end{equation}
where $L=2^b$ is the quantization level. We selected the bit depth, $b$, as 4, 6, and 8 kbps.
\vspace{-8pt}
\subsubsection{Optimization-based AF Attacks} \vspace{-6pt}
\label{sec:op_af}
Optimization-based AFs, such as FGSM~\cite{goodfellow2014explaining}, PGD~\cite{madry2017towards}, C{\&}W~\cite{carlini2017towards}, and DeepFool~\cite{moosavi2016deepfool}, introduce minimal perturbations to fool ADDs while preserving perceptual quality. Unlike statistical methods, these attacks iteratively optimize a loss to maximize misclassification.\\
% \textbf{FGSM}~\cite{goodfellow2014explaining, kawa2022defense} is a widely used AF attack in computer vision and image processing that generates perturbed samples in the direction of the sign of the gradient to maximize the wrong prediction. This method effectively deceives ADD models~\cite{zhang2020black} by subtly altering audio without changing the perceptual quality to listeners.
\textbf{FGSM}\cite{goodfellow2014explaining, kawa2022defense} is a popular AF attack in computer vision and image processing that perturbs inputs in the direction of the gradient sign to induce misclassification. It effectively fools ADDs\cite{zhang2020black} by subtly altering audio without affecting perceptual quality. Let $A$ be the input audio with label $y$. FGSM generates an attacked audio $A'$ using the loss $J(\theta, A, y)$ as follows:
\vspace{-8pt}
\begin{equation}
A' = A + \epsilon \cdot \text{sign} \left( \nabla_{A} J(\theta, A, y) \right)
\end{equation}
We set $\epsilon$ as 0.001, 0.05, 0.01, 0.1, and 0.2 to control the perturbation magnitude. \\
\textbf{PGD}~\cite{madry2017towards, kawa2022defense} is an iterative AF attack that applies small gradient-based perturbations at each step to project them within a bound to maintain imperceptibility. It produces stronger attacks than single-step methods, defined as:
\vspace{-8pt}
\begin{equation}
A_{t+1}' = 
\begin{cases}
A + \delta, & t = 0, \quad \delta \sim \mathcal{U}(-\epsilon, \epsilon) \\
\mathrm{Proj}_{\epsilon} \left( A_t' + \alpha \cdot \mathrm{sign} \left( \nabla_{A} J(\theta, A_t', y) \right) \right), & t \geq 0
\end{cases}
\end{equation}
where $A_t'$ and $\delta$ are the attacked sample at step $t$ and random noise. We set $\epsilon$ as 0.003, 0.007, 0.015, 0.03, and 0.06 with a step size $\alpha$ of 20 to control the perturbation magnitude.
\\
\textbf{C} \& \textbf{W}~\cite{carlini2017towards} is a powerful and widely used AF attack that crafts a targeted sample with minimal distortion. It formulates the process as a constrained optimization problem to find the smallest perturbation, defined as:
\vspace{-10pt}
\begin{equation}
\min_{\delta} \quad \| \delta \|_2^2 + c \cdot f(A + \delta), \quad
\text{where} \quad
f(A') = \max \left( \max_{i \neq y} Z_a(A') - Z_y(A'), -k \right)
\end{equation}
where $Z_a(.)$ and $Z_y(.)$ indicate the logits of attack and ground truth. We set confidence $c$ as 0, 10, 25, 35, and 50 to control perturbations and default $k$ as 0.\\
\textbf{DeepFool}~\cite{moosavi2016deepfool} is an optimization-based AF attack that iteratively perturbs inputs to cross the decision boundary with minimal distortion. Unlike FGSM and PGD, it computes the closest adversarial point using a linearized model approximation, defined as:
\vspace{-5pt}
\begin{align}
A' = A + \mathbf{r} = 
- \frac{Z_k(A) - Z_l(A)}
{\left\| \nabla Z_k(A) - \nabla Z_l(A) \right\|_2^2}
\cdot \left( \nabla Z_k(A) - \nabla Z_l(A) \right)
\end{align}
where $Z_k(A)$ and $Z_l(A)$ are the model’s confidence for classes $k$ and $l$. We set the step size to 50 and varied the overshoot in ${0.005, 0.01, 0.02, 0.03, 0.05}$ to balance the attack and quality.

\vspace{-15pt}
\section{{Results}}
\label{sec:results}

\vspace{-9pt}

\subsection{Experimental Setups} \vspace{-6pt}
This section details the environmental, training, and testing settings for overall evaluation.\\
\vspace{-25pt}
\subsubsection{Environmental Configurations} \vspace{-6pt}
All experiments were conducted on a Linux 24.04 system equipped with eight NVIDIA RTX 6000 Ada Generation GPUs, each with 49 GB of memory. We used Python 3.9 and PyTorch 2.2.2 as the software environment for implementing and evaluating the model.\\
\vspace{-25pt}
\subsubsection{Training and Testing Configurations} \vspace{-8pt}
We combined datasets $D1$, $D2$, and $D4$ for training, encompassing diverse generative methods, compression formats, and codecs. Testing used official splits from all five datasets ($D1$–$D5$) to improve generalization. The ADD models are trained for 50 epochs with a learning rate of 0.0001, batch size 256, using the Adam optimizer with weight decay 0.0001.\\
\vspace{-18pt}
\subsection{Performance Evaluations and Comparisons}  \vspace{-8pt}
\subsubsection{Baseline Detection Results}  \vspace{-8pt}
As discussed in Section~\ref{sec:dfd}, we evaluated two ADD groups: raw audio and spectrogram-based methods. Table~\ref{tab:baseline} lists the baseline detection results. In particular, spectrogram-based methods ($S1$-$S6$, achieving an average AUC and EER of 0.86 and 0.19) work better than raw methods ($R1$-$R6$, achieving an average AUC and EER of 0.75 and 0.30). Since all ADD models are trained on $D1$, $D2$, and $D4$, they generalize well to unseen datasets $D3$ and $D5$.
\vspace{-18pt}
\subsection{Results against Statistical AF Attacks} \vspace{-8pt}
We evaluated the resilience of ADDs against four statistical AFs: pitch shifting, filtering, noise addition, and quantization as described in Section~\ref{sec:st_af}. Table~\ref{tab:af_st} summarizes the average AUC and EER scores in five datasets.
For raw ADDs, performance degradation was substantial across all statistical AFs. Specifically, average AUC/EER dropped to 0.57/0.43 for pitch shifting, 0.59/0.42 for filtering, 0.52/0.47 for noise addition, and 0.60/0.42 for quantization. $R2$~\cite{wang2018csi}, $R3$~\cite{wu2020defense}, and $R4$~\cite{kawa2022defense} were notably vulnerable to filtering and noise, with AUCs falling below 0.30. In contrast, $R1$~\cite{jung2022pushing} and $R5$~\cite{hua2021towards} maintained relatively higher robustness due to enhanced temporal modeling and stronger feature representations.\\
On the other hand, spectrogram-based ADDs exhibited similar susceptibility to statistical AFs. As reported in Table~\ref{tab:af_st}, average AUC and EER were reduced to 0.50/0.44 for pitch shifting~\cite{chirkovskiy2025pitch}, 0.44/0.49 for filtering~\cite{wu2020defense}, 0.48/0.49 for noise addition~\cite{li2025measuring}, and 0.43/0.53 for quantization~\cite{grama2015quantization}. Despite strong visual learning, spectrogram-based methods show instability under simple AFs, emphasizing the need for more robust ADD methods.
\vspace{-18pt}

\begin{table*}[!b]
\centering
\caption{Baseline Performance of ADD Methods.}
\vspace{5pt}
\begin{adjustbox}{width=\textwidth}
\begin{tabular}{lcccccc|lcccccc}
\toprule
\multicolumn{7}{l|}{\textbf{Raw Audio-based ADD Methods}} & \multicolumn{7}{l}{\textbf{Spectrogram-based ADD Methods}} \\
\midrule
Method & \textbf{D1~\cite{todisco2019asvspoof}} & \textbf{D2~\cite{liu2023asvspoof}} & \textbf{D3~\cite{wang2024asvspoof}} & \textbf{D4~\cite{xie2025codecfake}} & \textbf{D5~\cite{frank2021wavefake}} & \textbf{Avg.} &
Method & \textbf{D1~\cite{todisco2019asvspoof}} & \textbf{D2~\cite{liu2023asvspoof}} & \textbf{D3~\cite{wang2024asvspoof}} & \textbf{D4~\cite{xie2025codecfake}} & \textbf{D5~\cite{frank2021wavefake}} & \textbf{Avg.} \\
\midrule
$R1$~\cite{jung2022pushing} & 0.99 / 0.05 & 0.73 / 0.32 & 0.68 / 0.36  & 0.83 / 0.25  &  0.96 / 0.10 & 0.83 / 0.21 &
$S1$~\cite{wani2024abc} & 0.99 / 0.04 & 0.78 / 0.27 & 0.57 / 0.45 & 0.93 / 0.13 & 0.89 / 0.19 & 0.83 / 0.22 \\
$R2$~\cite{wang2018csi} & 0.98 / 0.07 & 0.80 / 0.24 & 0.53 / 0.46  & 0.66 / 0.40 & 0.56 / 0.47 & 0.71 / 0.33  &
$S2$~\cite{tan2019efficientnet} & 0.99 / 0.04 & 0.83 / 0.26 & 0.58 / 0.42 & 0.93 / 0.14 & 0.81 / 0.25 & 0.83 / 0.21 \\
$R3$~\cite{wu2020defense} & 0.93 / 0.16 & 0.78 / 0.28 & 0.58 / 0.45  & 0.87 / 0.23 & 0.54 / 0.49 & 0.74 / 0.32 &
$S3$~\cite{dang2020detection} & 1.00 / 0.04 & 0.85 / 0.25 & 0.64 / 0.40 & 0.99 / 0.06 & 0.99 / 0.04 & 0.89 / 0.16 \\
$R4$~\cite{kawa2022defense} & 0.98 / 0.21 & 0.85 / 0.22 & 0.54 / 0.48  & 0.72 / 0.33 & 0.54 / 0.48 & 0.73 / 0.34 &
$S4$~\cite{ni2022core} & 0.98 / 0.09 & 0.82 / 0.26 & 0.65 / 0.40 & 0.99 / 0.07 & 0.98 / 0.08 & 0.88 / 0.18 \\
$R5$~\cite{hua2021towards} & 0.98 / 0.06 & 0.75 / 0.31 & 0.65 / 0.39  &  0.86 / 0.23 & 0.67 / 0.38 & 0.78 / 0.27 &
$S5$~\cite{qian2020thinking} & 0.99 / 0.04 & 0.79 / 0.29 & 0.63 / 0.40 & 0.97 / 0.08 & 0.96 / 0.11 & 0.87 / 0.18 \\
$R6$~\cite{hua2021towards} & 0.99 / 0.05 & 0.76 / 0.30 & 0.52 / 0.48  &  0.75 / 0.34 & 0.61 / 0.42 & 0.73 / 0.32 &
$S6$~\cite{ziabary2021countermeasure} & 0.98 / 0.06 & 0.70 / 0.36 & 0.66 / 0.40 & 0.97 / 0.10 & 0.98 / 0.07 & 0.86 / 0.20 \\
\midrule
\textbf{Avg.} & \textbf{0.98 / 0.10} & \textbf{0.78 / 0.28} & \textbf{0.58 / 0.43} & \textbf{0.78 / 0.30} & \textbf{0.65 / 0.39} & \textbf{0.75 / 0.30} &
\textbf{Avg.} & \textbf{0.99 / 0.05} & \textbf{0.80 / 0.28} & \textbf{0.62 / 0.41} & \textbf{0.96 / 0.10} & \textbf{0.94 / 0.12} & \textbf{0.86 / 0.19} \\
\bottomrule
\end{tabular}
\end{adjustbox}
\label{tab:baseline}
\vspace{-13pt}
\end{table*}

\begin{table*}[!b]
\centering
\caption{Performance of Statistical AF Attacks on ADD Methods.} 

\vspace{5pt}
\begin{adjustbox}{width=\textwidth}
\begin{tabular}{lcccccc|lcccccc}
\toprule
\multicolumn{7}{l|}{\textbf{Raw Audio-based ADD Methods}} & \multicolumn{7}{l}{\textbf{Spectrogram-based ADD Methods}} \\
\midrule
Method & \textbf{D1~\cite{todisco2019asvspoof}} & \textbf{D2~\cite{liu2023asvspoof}} & \textbf{D3~\cite{wang2024asvspoof}} & \textbf{D4~\cite{xie2025codecfake}} & \textbf{D5~\cite{frank2021wavefake}} & \textbf{Avg.} &
Method & \textbf{D1~\cite{todisco2019asvspoof}} & \textbf{D2~\cite{liu2023asvspoof}} & \textbf{D3~\cite{wang2024asvspoof}} & \textbf{D4~\cite{xie2025codecfake}} & \textbf{D5~\cite{frank2021wavefake}} & \textbf{Avg.} \\

\midrule
\multicolumn{7}{l|}{\textbf{Pitch Shifting~\cite{chirkovskiy2025pitch}}} & \multicolumn{7}{l}{\textbf{Pitch Shifting~\cite{chirkovskiy2025pitch}}} \\
\midrule
$R1$~\cite{jung2022pushing}     & 0.92 / .12 & 0.50 / 0.51 & 0.43 / 0.56 & 0.75 / 0.30 & 0.75 / 0.30 &  0.67 / 0.36 & $S1$~\cite{wani2024abc}  & 0.31 / 0.84 & 0.54 / 0.23 & 0.68 / 0.32 & 0.31 / 0.67 & 0.56 / 0.23 & 0.48 / 0.46\\
$R2$~\cite{wang2018csi}     & 0.86 / 0.21 & 0.56 / 0.42 & 0.30 / 0.65  & 0.54 / 0.44 & 0.29 / 0.63 & 0.51 / 0.47 & $S2$~\cite{tan2019efficientnet}  & 0.37 / 0.63 & 0.40 / 0.60 & 0.37 / 0.63 & 0.56 / 0.44 & 0.41 / 0.58 & 0.42 / 0.58 \\
$R3$~\cite{wu2020defense}     & 0.92 / 0.13 & 0.60 / 0.41 & 0.19 / 0.74   & 0.80 / 0.27 & 0.15 / 0.78 & 0.53 / 0.47 & $S3$~\cite{dang2020detection} & 0.35 / 0.83 & 0.51 / 0.25 & 0.66 / 0.35 & 0.59 / 0.39 & 0.32 / 0.84 & 0.49 / 0.53 \\
$R4$~\cite{kawa2022defense}     & 0.80 / 0.25 & 0.81 / 0.23 & 0.36 / 0.60    & 0.51 / 0.50 & 0.12 / 0.77 & 0.52 / 0.47 & $S4$~\cite{ni2022core} & 0.47 / 0.76 & 0.44 / 0.78 & 0.69 / 0.30 & 0.59 / 0.41 & 0.37 / 0.80 & 0.51 / 0.61 \\
$R5$~\cite{hua2021towards}     & 0.95 / 0.12 & 0.70 / 0.36 & 0.30 / 0.65   & 0.84 / 0.22 & 0.39 / 0.56 & 0.64 / 0.38 & $S5$~\cite{qian2020thinking} & 0.58 / 0.21 & 0.53 / 0.24 & 0.54 / 0.46 & 0.33 / 0.66 & 0.53 / 0.25 & 0.50 / 0.36 \\
$R6$~\cite{hua2021towards}     & 0.93 / 0.14 & 0.41 / 0.52 & 0.23 / 0.69   & 0.78 / 0.28 & 0.33 / 0.61 & 0.54 / 0.45 & $S6$~\cite{ziabary2021countermeasure} & 0.54 / 0.22 & 0.50 / 0.26 & 0.96 / 0.11 & 0.34 / 0.68 & 0.51 / 0.26 & 0.57 / 0.31 \\
\midrule
\textbf{Avg.} & \textbf{0.90 / 0.16} & \textbf{0.60 / 0.41} & \textbf{0.30 / 0.65} & \textbf{0.70 / 0.34} & \textbf{0.34 / 0.61} & \textbf{0.57 / 0.43} &
\textbf{Avg.} & \textbf{0.45 / 0.45} & \textbf{0.49 / 0.39} & \textbf{0.64 / 0.36} & \textbf{0.45 / 0.54} & \textbf{0.45 / 0.49} & \textbf{0.50 / 0.44} \\
\midrule
\multicolumn{7}{l|}{\textbf{Filtering~\cite{wu2020defense}}} & \multicolumn{7}{l}{\textbf{Filtering~\cite{wu2020defense}}} \\
\midrule
$R1$~\cite{jung2022pushing}     & 0.98 / 0.03 & 0.88 / 0.15 & 0.73 / 0.28  & 0.96 / 0.10  & 0.98 / 0.04 & 0.91 / 0.12 & $S1$~\cite{wani2024abc}  & 0.38 / 0.81 & 0.58 / 0.21 & 0.30 / 0.70 & 0.35 / 0.65 & 0.43 / 0.78 & 0.41 / 0.63 \\
$R2$~\cite{wang2018csi}     & 0.92 / 0.15 & 0.48 / 0.49 &  0.21 / 0.73  & 0.82 / 0.25 & 0.52 / 0.49 & 0.59 / 0.42 & $S2$~\cite{tan2019efficientnet}  & 0.35 / 0.65 & 0.46 / 0.53 & 0.47 / 0.53 & 0.31 / 0.69 & 0.53 / 0.45 & 0.42 / 0.57 \\
$R3$~\cite{wu2020defense}     & 0.34 / 0.60 & 0.14 / 0.75 & 0.03 / 0.90 & 0.52 / 0.47 & 0.14 / 0.78 & 0.23 / 0.70 & $S3$~\cite{dang2020detection} & 0.35 / 0.83 & 0.49 / 0.26 & 0.27 / 0.74 & 0.36 / 0.65 & 0.31 / 0.84 & 0.36 / 0.66 \\
$R4$~\cite{kawa2022defense}     & 0.35 / 0.62 & 0.11 / 0.82 & 0.03 / 0.94 & 0.10 / 0.83 & 0.03 / 0.97 & 0.12 / 0.84  & $S4$~\cite{ni2022core} & 0.60 / 0.20 & 0.31 / 0.84 & 0.35 / 0.65 & 0.47 / 0.52 & 0.57 / 0.23 & 0.46 / 0.49 \\
$R5$~\cite{hua2021towards}     & 0.97 / 0.08 & 0.81 / 0.24 & 0.51 / 0.48 & 0.95 / 0.11 & 0.86 / 0.21 &  0.82 / 0.22 & $S5$~\cite{qian2020thinking} & 0.60 / 0.20 & 0.58 / 0.22 & 0.40 / 0.59 & 0.47 / 0.53 & 0.56 / 0.23 & 0.52 / 0.35 \\ 
$R6$~\cite{hua2021towards}     & 0.99 / 0.01 & 0.77 / 0.25 & 0.69 / 0.35  &  0.98 / 0.06 & 0.76 / 0.33 & 0.84 / 0.20 & $S6$~\cite{ziabary2021countermeasure} & 0.56 / 0.23 & 0.54 / 0.24 & 0.30 / 0.69 & 0.38 / 0.62 & 0.52 / 0.24 & 0.46 / 0.40 \\
\midrule
\textbf{Avg.} & \textbf{0.76 / 0.25} & \textbf{0.53 / 0.45} & \textbf{0.37 / 0.61} & \textbf{0.72 / 0.30} & \textbf{0.55 / 0.47} & \textbf{0.59 / 0.42} & \textbf{Avg.} & \textbf{0.47 / 0.35} & \textbf{0.48 / 0.38} & \textbf{0.35 / 0.65} & \textbf{0.39 / 0.61} & \textbf{0.49 / 0.46} & \textbf{0.44 / 0.49} \\
\midrule
\multicolumn{7}{l|}{\textbf{Noise Addition~\cite{li2025measuring}}} & \multicolumn{7}{l}{\textbf{Noise Addition~\cite{li2025measuring}}} \\
\midrule
$R1$~\cite{jung2022pushing}     & 0.97 / 0.04 & 0.71 / 0.32 & 0.46 / 0.55   & 0.94 / 0.10 & 0.97 / 0.07 & 0.81 / 0.22 & $S1$~\cite{wani2024abc}  & 0.49 / 0.75 & 0.41 / 0.79 & 0.48 / 0.53 & 0.52 / 0.49 & 0.38 / 0.81 & 0.46 / 0.67 \\
$R2$~\cite{wang2018csi}     & 0.64 / 0.40 & 0.32 / 0.62 & 0.25 / 0.66   & 0.10 / 0.88 & 0.13 / 0.79 & 0.29 / 0.67 & $S2$~\cite{tan2019efficientnet}  & 0.37 / 0.63 & 0.51 / 0.50 & 0.34 / 0.66 & 0.54 / 0.46 & 0.41 / 0.58 & 0.43 / 0.57 \\
$R3$~\cite{wu2020defense}     & 0.74 / 0.32 & 0.37 / 0.60 & 0.14 / 0.80  & 0.76 / 0.31  & 0.20 / 0.74 & 0.44 / 0.55 & $S3$~\cite{dang2020detection} & 0.46 / 0.77 & 0.48 / 0.76 & 0.21 / 0.80 & 0.54 / 0.46 & 0.34 / 0.83 & 0.41 / 0.72 \\
$R4$~\cite{kawa2022defense}     & 0.73 / 0.28 & 0.11 / 0.82 & 0.09 / 0.85  & 0.23 / 0.71 & 0.04 / 0.87 & 0.24 / 0.71 & $S4$~\cite{ni2022core} & 0.57 / 0.22 & 0.59 / 0.21 & 0.67 / 0.32 & 0.54 / 0.47 & 0.50 / 0.26 & 0.57 / 0.30 \\
$R5$~\cite{hua2021towards}     & 0.89 / 0.17 & 0.81 / 0.24 & 0.19 / 0.76   & 0.92 / 0.16 & 0.66 / 0.40 & 0.69 / 0.35 & $S5$~\cite{qian2020thinking} & 0.37 / 0.81 & 0.53 / 0.24 & 0.59 / 0.41 & 0.39 / 0.61 & 0.55 / 0.23 & 0.49 / 0.46 \\
$R6$~\cite{hua2021towards}     & 0.95 / 0.10 & 0.77 / 0.25 & 0.24 / 0.70 & 0.85 / 0.24 & 0.47 / 0.52 & 0.66 / 0.36 & $S6$~\cite{ziabary2021countermeasure} & 0.55 / 0.24 & 0.52 / 0.25 & 0.54 / 0.46 & 0.45 / 0.55 & 0.53 / 0.23 & 0.52 / 0.35 \\
\midrule
\textbf{Avg.} &\textbf{ 0.77 / 0.22} &\textbf{ 0.51 / 0.47} & \textbf{0.23 / 0.71 }& \textbf{0.63 / 0.40} & \textbf{0.41 / 0.56} & \textbf{0.52 / 0.47}  & 
\textbf{Avg.} & \textbf{0.46 / 0.44} & \textbf{0.52 / 0.46} & \textbf{0.46 / 0.53} & \textbf{0.50 / 0.51} & \textbf{0.45 / 0.49} & \textbf{0.48 / 0.49} \\
\midrule
\multicolumn{7}{l|}{\textbf{Quantization~\cite{grama2015quantization}}} & \multicolumn{7}{l}{\textbf{Quantization~\cite{grama2015quantization}}} \\
\midrule
$R1$~\cite{jung2022pushing}     & 0.88 / 0.20 & 0.56 / 0.47 &  0.46 / 0.49 & 0.68 / 0.39 & 0.64 / 0.39 & 0.64 / 0.39 & $S1$~\cite{wani2024abc}  & 0.37 / 0.82 & 0.47 / 0.76 & 0.24 / 0.76 & 0.52 / 0.48 & 0.31 / 0.84 & 0.38 / 0.73 \\
$R2$~\cite{wang2018csi}     & 0.92 / 0.23 & 0.49 / 0.46 & 0.30 / 0.63   & 0.59 / 0.42  & 0.34 / 0.60 & 0.53 / 0.47 & $S2$~\cite{tan2019efficientnet}  & 0.49 / 0.51 & 0.41 / 0.60 & 0.52 / 0.48 & 0.34 / 0.66 & 0.45 / 0.53 & 0.44 / 0.56 \\
$R3$~\cite{wu2020defense}     & 0.85 / 0.23 & 0.66 / 0.36 & 0.28 / 0.66  &  0.40 / 0.58  & 0.04 / 0.89 & 0.45 / 0.54 & $S3$~\cite{dang2020detection} & 0.56 / 0.22 & 0.34 / 0.83 & 0.50 / 0.50 & 0.28 / 0.73 & 0.48 / 0.76 & 0.43 / 0.61 \\
$R4$~\cite{kawa2022defense}     & 0.83 / 0.22 & 0.68 / 0.33 & 0.14 / 0.79 & 0.53 / 0.48   & 0.06 / 0.85 & 0.45 / 0.53 & $S4$~\cite{ni2022core} & 0.56 / 0.22 & 0.60 / 0.20 & 0.47 / 0.51 & 0.42 / 0.57 & 0.35 / 0.81 & 0.48 / 0.46 \\
$R5$~\cite{hua2021towards}     & 0.99 / 0.03 & 0.79 / 0.28 & 0.69 / 0.35   & 0.85 / 0.23   & 0.69 / 0.37 & 0.80 / 0.25 & $S5$~\cite{qian2020thinking} & 0.35 / 0.83 & 0.39 / 0.80 & 0.44 / 0.57 & 0.31 / 0.69 & 0.38 / 0.80 & 0.37 / 0.74 \\
$R6$~\cite{hua2021towards}     & 0.99 / 0.04 & 0.74 / 0.30 & 0.49 / 0.51  & 0.82 / 0.28  & 0.54 / 0.47  &  0.72 / 0.32 & $S6$~\cite{ziabary2021countermeasure} & 0.53 / 0.25 & 0.56 / 0.22 & 0.33 / 0.67 & 0.36 / 0.64 & 0.49 / 0.25 & 0.45 / 0.40 \\
\midrule
\textbf{Avg.} & \textbf{0.91 / 0.15} & \textbf{0.65 / 0.37} & \textbf{0.39 / 0.57} & \textbf{0.64 / 0.40} & \textbf{0.38 / 0.59} & \textbf{0.60 / 0.42} & 
\textbf{Avg.} & \textbf{0.48 / 0.44} & \textbf{0.46 / 0.57} & \textbf{0.42 / 0.57} & \textbf{0.37 / 0.62} & \textbf{0.41 / 0.66} & \textbf{0.43 / 0.53} \\
\bottomrule
\end{tabular}
\end{adjustbox}
\label{tab:af_st}
\vspace{-10pt}
\end{table*}

\begin{table*}[t]
\centering
\caption{Performance of Optimization-based AF Attacks on ADD Methods.} 
\vspace{5pt}
\begin{adjustbox}{width=\textwidth}
\begin{tabular}{lcccccc|lcccccc}
\toprule
\multicolumn{7}{l|}{\textbf{Raw Audio-based ADD Methods}} & \multicolumn{7}{l}{\textbf{Spectrogram-based ADD Methods}} \\
\midrule
Method & \textbf{D1~\cite{todisco2019asvspoof}} & \textbf{D2~\cite{liu2023asvspoof}} & \textbf{D3~\cite{wang2024asvspoof}} & \textbf{D4~\cite{xie2025codecfake}} & \textbf{D5~\cite{frank2021wavefake}} & \textbf{Avg.} &
Method & \textbf{D1~\cite{todisco2019asvspoof}} & \textbf{D2~\cite{liu2023asvspoof}} & \textbf{D3~\cite{wang2024asvspoof}} & \textbf{D4~\cite{xie2025codecfake}} & \textbf{D5~\cite{frank2021wavefake}} & \textbf{Avg.} \\
\midrule
\multicolumn{7}{l|}{\textbf{FGSM~\cite{goodfellow2014explaining}}} & \multicolumn{7}{l}{\textbf{FGSM~\cite{goodfellow2014explaining}}} \\
\midrule
$R1$~\cite{jung2022pushing}     & 0.69 / 0.37 & 0.33 / 0.65 & 0.17 / 0.78 & 0.69 / 0.36 & 0.77 / 0.31 & 0.53 / 0.50 & $S1$~\cite{wani2024abc} & 0.52 / 0.24 & 0.40 / 0.80 & 0.21 / 0.85 & 0.55 / 0.44 & 0.30 / 0.62 & 0.40 / 0.59 \\
$R2$~\cite{wang2018csi}         & 0.21 / 0.73 & 0.03 / 0.94 & 0.01 / 0.97 & 0.02 / 0.97 & 0.00 / 1.00 & 0.05/0.92 & $S2$~\cite{tan2019efficientnet} & 0.47 / 0.52 & 0.49 / 0.50 & 0.23 / 0.63 & 0.54 / 0.46 & 0.44 / 0.53 & 0.43 / 0.53 \\
$R3$~\cite{wu2020defense}       & 0.69 / 0.39 & 0.24 / 0.67 & 0.21 / 0.68 & 0.64 / 0.38 & 0.17 / 0.82 & 0.39 / 0.59 & $S3$~\cite{dang2020detection} & 0.35 / 0.82 & 0.22 / 0.89 & 0.21 / 0.74 & 0.42 / 0.58 & 0.27 / 0.86 & 0.29 / 0.78 \\
$R4$~\cite{kawa2022defense}     & 0.59 / 0.38 & 0.30 / 0.59 & 0.13 / 0.75 & 0.32 / 0.59 & 0.09 / 0.78 & 0.28 / 0.62 & $S4$~\cite{ni2022core} & 0.33 / 0.83 & 0.33 / 0.53 & 0.25 / 0.50 & 0.47 / 0.57 & 0.22 / 0.89 & 0.32 / 0.67 \\
$R5$~\cite{hua2021towards}      & 0.88 / 0.20 & 0.58 / 0.44 & 0.20 / 0.72 & 0.91 / 0.17 & 0.51 / 0.48 & 0.62 / 0.40 & $S5$~\cite{qian2020thinking} & 0.31 / 0.85 & 0.52 / 0.25 & 0.25 / 0.50 & 0.26 / 0.74 & 0.53 / 0.24 & 0.37 / 0.52 \\
$R6$~\cite{hua2021towards}      & 0.70 / 0.34 & 0.16 / 0.75 & 0.01 / 0.94 & 0.29 / 0.64 & 0.05 / 0.87 & 0.24 / 0.71 & $S6$~\cite{ziabary2021countermeasure} & 0.40 / 0.65 & 0.39 / 0.60 & 0.23 / 0.64 & 0.45 / 0.56 & 0.35 / 0.63 & 0.37 / 0.62 \\
\midrule
\textbf{Avg.} & \textbf{0.63 / 0.40} & \textbf{0.27 / 0.67} & \textbf{0.12 / 0.81} & \textbf{0.48 / 0.52} & \textbf{0.26 / 0.71} & \textbf{0.35 / 0.62} & \textbf{Avg.} & \textbf{0.40 / 0.65} & \textbf{0.39 / 0.59} & \textbf{0.23 / 0.64} & \textbf{0.45 / 0.56} & \textbf{0.35 / 0.63} & \textbf{0.36 / 0.61} \\
\midrule
\multicolumn{7}{l|}{\textbf{PGD~\cite{madry2017towards}}} & \multicolumn{7}{l}{\textbf{PGD~\cite{madry2017towards}}} \\
\midrule
$R1$~\cite{jung2022pushing} & 0.41 / 0.60 & 0.25 / 0.69 & 0.12 / 0.82 & 0.23 / 0.71 & 0.43 / 0.57 & 0.29 / 0.68 & $S1$~\cite{wani2024abc} & 0.51 / 0.24 & 0.34 / 0.83 & 0.21 / 0.85 & 0.41 / 0.58 & 0.34 / 0.60 & 0.36 / 0.62 \\
$R2$~\cite{wang2018csi}     & 0.09 / 0.87 & 0.01 / 0.98 & 0.10 / 0.90 & 0.04 / 0.96 & 0.03 / 0.97 & 0.05 / 0.94 & $S2$~\cite{tan2019efficientnet} & 0.35 / 0.65 & 0.43 / 0.54 & 0.23 / 0.63 & 0.33 / 0.67 & 0.27 / 0.74 & 0.32 / 0.65 \\
$R3$~\cite{wu2020defense}   & 0.03 / 0.97 & 0.02 / 0.93 & 0.05 / 0.95 & 0.03 / 0.93 & 0.07 / 0.93 & 0.04 / 0.94 & $S3$~\cite{dang2020detection} & 0.57 / 0.22 & 0.24 / 0.88 & 0.21 / 0.74 & 0.28 / 0.72 & 0.54 / 0.23 & 0.37 / 0.56 \\
$R4$~\cite{kawa2022defense} & 0.20 / 0.68 & 0.10 / 0.80 & 0.04 / 0.96 & 0.02 / 0.95 & 0.09 / 0.91 & 0.09 / 0.86 & $S4$~\cite{ni2022core} & 0.34 / 0.83 & 0.27 / 0.56 & 0.25 / 0.50 & 0.56 / 0.44 & 0.37 / 0.81 & 0.36 / 0.63 \\
$R5$~\cite{hua2021towards}  & 0.07 / 0.85 & 0.02 / 0.91 & 0.11 / 0.89 & 0.02 / 0.93 & 0.04 / 0.96 & 0.05 / 0.91 & $S5$~\cite{qian2020thinking} & 0.40 / 0.80 & 0.30 / 0.84 & 0.25 / 0.50 & 0.23 / 0.76 & 0.52 / 0.25 & 0.34 / 0.63 \\
$R6$~\cite{hua2021towards}  & 0.10 / 0.82 & 0.01 / 0.98 & 0.09 / 0.91 & 0.01 / 0.99 & 0.05 / 0.95 & 0.05 / 0.93 & $S6$~\cite{ziabary2021countermeasure}  & 0.43 / 0.55 & 0.32 / 0.73 & 0.23 / 0.64 & 0.36 / 0.63 & 0.41 / 0.53 & 0.35 / 0.62 \\
\midrule
\textbf{Avg.} & \textbf{0.15 / 0.80} & \textbf{0.07 / 0.88} & \textbf{0.08 / 0.91} & \textbf{0.06 / 0.91} & \textbf{0.12 / 0.88} &  \textbf{0.09 / 0.87} & \textbf{Avg.} & \textbf{0.43 / 0.55} & \textbf{0.32 / 0.73} & \textbf{0.23 / 0.64} & \textbf{0.36 / 0.63} & \textbf{0.41 / 0.53} & \textbf{0.35 / 0.62} \\
\midrule
\multicolumn{7}{l|}{\textbf{C \& W~\cite{carlini2017towards}}} & \multicolumn{7}{l}{\textbf{C \& W~\cite{carlini2017towards}}} \\
\midrule
$R1$~\cite{jung2022pushing} & 0.98 / 0.08 & 0.81 / 0.23 & 0.55 / 0.41 & 0.62 / 0.43 & 0.94 / 0.11 & 0.78 / 0.25 & $S1$~\cite{wani2024abc} & 0.57 / 0.21 & 0.29 / 0.85 & 0.21 / 0.85 & 0.53 / 0.47 & 0.38 / 0.59 & 0.40 / 0.59 \\
$R2$~\cite{wang2018csi}     & 0.85 / 0.23 & 0.26 / 0.66 & 0.17 / 0.78 & 0.44 / 0.54 & 0.33 / 0.61 & 0.41 / 0.56 & $S2$~\cite{tan2019efficientnet} & 0.23 / 0.77 & 0.55 / 0.43 & 0.23 / 0.63 & 0.33 / 0.67 & 0.21 / 0.78 & 0.31 / 0.66 \\
$R3$~\cite{wu2020defense}   & 0.93 / 0.16 & 0.78 / 0.29 & 0.49 / 0.49 & 0.71 / 0.35 & 0.30 / 0.68 & 0.64 / 0.39 & $S3$~\cite{dang2020detection} & 0.50 / 0.25 & 0.36 / 0.81 & 0.21 / 0.74 & 0.41 / 0.59 & 0.50 / 0.26 & 0.40 / 0.53 \\
$R4$~\cite{kawa2022defense} & 0.81 / 0.23 & 0.73 / 0.28 & 0.23 / 0.74 & 0.42 / 0.54 & 0.05 / 0.87 & 0.45 / 0.53 & $S4$~\cite{ni2022core} & 0.20 / 0.90 & 0.50 / 0.44 & 0.25 / 0.50 & 0.43 / 0.59 & 0.47 / 0.76 & 0.37 / 0.64 \\
$R5$~\cite{hua2021towards}  & 0.85 / 0.23 & 0.51 / 0.49 & 0.12 / 0.81 & 0.63 / 0.41 & 0.24 / 0.70 & 0.47 / 0.51 & $S5$~\cite{qian2020thinking} & 0.58 / 0.21 & 0.33 / 0.82 & 0.25 / 0.50 & 0.40 / 0.62 & 0.27 / 0.86 & 0.37 / 0.60 \\
$R6$~\cite{hua2021towards}  & 0.82 / 0.25 & 0.27 / 0.66 & 0.02 / 0.95 & 0.18 / 0.75 & 0.06 / 0.86 & 0.27 / 0.69 & $S6$~\cite{ziabary2021countermeasure}  & 0.42 / 0.47 & 0.41 / 0.67 & 0.23 / 0.64 & 0.42 / 0.59 & 0.37 / 0.65 & 0.37 / 0.60 \\
\midrule
\textbf{Avg} & \textbf{0.87 / 0.20} & \textbf{0.56 / 0.43} & \textbf{0.26 / 0.70} & \textbf{0.51 / 0.49} & \textbf{0.32 / 0.64} & \textbf{0.51 / 0.47} & \textbf{Avg.} & \textbf{0.42 / 0.47} & \textbf{0.41 / 0.67} & \textbf{0.23 / 0.64} & \textbf{0.42 / 0.59} & \textbf{0.37 / 0.65} & \textbf{0.37 / 0.60} \\
\midrule
\multicolumn{7}{l|}{\textbf{DeepFool~\cite{moosavi2016deepfool}}} & \multicolumn{7}{l}{\textbf{DeepFool~\cite{moosavi2016deepfool}}} \\
\midrule
$R1$~\cite{jung2022pushing}     & 0.98 / 0.07 & 0.70 / 0.33 & 0.55 / 0.41 & 0.76 / 0.30 & 0.96 / 0.11 & 0.79 / 0.24 & $S1$~\cite{wani2024abc} & 0.56 / 0.22 & 0.26 / 0.87 & 0.21 / 0.85 & 0.41 / 0.58 & 0.27 / 0.64 & 0.34 / 0.63 \\
$R2$~\cite{wang2018csi}         & 0.96 / 0.12 & 0.63 / 0.31 & 0.21 / 0.74 & 0.62 / 0.42 & 0.48 / 0.50 & 0.57 / 0.40 & $S2$~\cite{tan2019efficientnet} & 0.22 / 0.78 & 0.51 / 0.46 & 0.23 / 0.63 & 0.53 / 0.47 & 0.21 / 0.78 & 0.34 / 0.62 \\
$R3$~\cite{wu2020defense}       & 0.91 / 0.17 & 0.54 / 0.43 & 0.14 / 0.83 & 0.69 / 0.37 & 0.25 / 0.66 & 0.50 / 0.39 & $S3$~\cite{dang2020detection} & 0.38 / 0.81 & 0.51 / 0.25 & 0.21 / 0.74 & 0.26 / 0.74 & 0.55 / 0.23 & 0.38 / 0.55 \\
$R4$~\cite{kawa2022defense}     & 0.79 / 0.24 & 0.72 / 0.28 & 0.23 / 0.71 & 0.37 / 0.56 & 0.12 / 0.77 & 0.41 / 0.49 & $S4$~\cite{ni2022core} & 0.58 / 0.21 & 0.54 / 0.42 & 0.25 / 0.50 & 0.41 / 0.60 & 0.32 / 0.84 & 0.42 / 0.51 \\
$R5$~\cite{hua2021towards}      & 0.90 / 0.18 & 0.57 / 0.44 & 0.15 / 0.76 & 0.80 / 0.26 & 0.38 / 0.58 & 0.64 / 0.34 & $S5$~\cite{qian2020thinking} & 0.30 / 0.85 & 0.48 / 0.28 & 0.25 / 0.50 & 0.37 / 0.63 & 0.28 / 0.85 & 0.34 / 0.62 \\
$R6$~\cite{hua2021towards}      & 0.88 / 0.19 & 0.33 / 0.53 & 0.03 / 0.92 & 0.44 / 0.52 & 0.10 / 0.78 & 0.44 / 0.46 & $S6$~\cite{ziabary2021countermeasure}  & 0.41 / 0.57 & 0.46 / 0.38 & 0.23 / 0.64 & 0.40 / 0.60 & 0.33 / 0.67 & 0.37 / 0.57 \\
\midrule
\textbf{Avg}& \textbf{0.90 / 0.16 }& \textbf{0.68 / 0.34} & \textbf{0.24 / 0.70} & \textbf{0.61 / 0.35} & \textbf{0.38 / 0.37} & \textbf{0.55 / 0.47} & \textbf{Avg.} & \textbf{0.41 / 0.57} & \textbf{0.46 / 0.38} & \textbf{0.23 / 0.64} & \textbf{0.40 / 0.60} & \textbf{0.33 / 0.67} & \textbf{0.37 / 0.57} \\
\bottomrule
\end{tabular}
\end{adjustbox}
\label{tab:opt_af}
\vspace{-10pt}
\end{table*}

\begin{table*}[t]
\centering
\caption{Performance of Defense against AF Attacks.} 
\vspace{5pt}
\begin{adjustbox}{width=\textwidth}
\begin{tabular}{lcccccc|lcccccc}
\toprule
\multicolumn{7}{l|}{\textbf{Raw Audio-based ADD Methods}} & \multicolumn{7}{l}{\textbf{Spectrogram-based ADD Methods}} \\
\midrule
Method & \textbf{D1~\cite{todisco2019asvspoof}} & \textbf{D2~\cite{liu2023asvspoof}} & \textbf{D3~\cite{wang2024asvspoof}} & \textbf{D4~\cite{xie2025codecfake}} & \textbf{D5~\cite{frank2021wavefake}} & \textbf{Avg.} &
Method & \textbf{D1~\cite{todisco2019asvspoof}} & \textbf{D2~\cite{liu2023asvspoof}} & \textbf{D3~\cite{wang2024asvspoof}} & \textbf{D4~\cite{xie2025codecfake}} & \textbf{D5~\cite{frank2021wavefake}} & \textbf{Avg.} \\
\midrule
$R1$~\cite{jung2022pushing} & 0.95 / 0.06 & 0.54 / 0.41 & 0.74 / 0.35 & 0.91 / 0.10 & 0.87 / 0.19 & 0.80 / 0.22 & $S3$~\cite{dang2020detection}  & 0.99 / 0.06 & 0.83 / 0.11 & 0.61 / 0.42 & 0.99 / 0.01 & 0.82 / 0.25 & 0.85 / 0.17 \\
$R5$~\cite{hua2021towards} & 0.89 / 0.13 & 0.67 / 0.38 & 0.58 / 0.43 & 0.78 / 0.31 & 0.61 / 0.44 & 0.71 / 0.34 & $S4$~\cite{ni2022core} & 0.98 / 0.07 & 0.79 / 0.12 & 0.59 / 0.45 & 0.99 / 0.01 & 0.75 / 0.30 & 0.82 / 0.19 \\
\midrule
\textbf{Avg} & \textbf{0.92 / 0.10} & \textbf{0.61 / 0.40} & \textbf{0.66 / 0.39} & \textbf{0.85 / 0.21} & \textbf{0.74 / 0.32} & \textbf{0.76 / 0.28} & 
\textbf{Avg}  & \textbf{0.99 / 0.06} & \textbf{0.81 / 0.12} & \textbf{0.60 / 0.44} & \textbf{0.99 / 0.01} & \textbf{0.79 / 0.28} & \textbf{0.83 / 0.18} \\
\bottomrule
\end{tabular}
\end{adjustbox}
\label{tab:defense}
\vspace{-12pt}
\end{table*}

\begin{table}[!t]
\caption{Qualitative Results of non-Attacked and Attacked Samples.}
\vspace{5pt}
\begin{adjustbox}{width=\textwidth}
\begin{tabular}{lcccccc|lcccccc}
\toprule
\multicolumn{7}{l|}{\textbf{Raw Audio-based ADD Methods}} & \multicolumn{7}{l}{\textbf{Spectrogram-based ADD Methods}} \\
\midrule
\textbf{Method} & \textbf{Metric} & \textbf{D1~\cite{todisco2019asvspoof}} & \textbf{D2~\cite{liu2023asvspoof}} & \textbf{D3~\cite{wang2024asvspoof}} & \textbf{D4~\cite{xie2025codecfake}} & \textbf{D5~\cite{frank2021wavefake}} &
\textbf{Method} & \textbf{Metric}  & \textbf{D1~\cite{todisco2019asvspoof}} & \textbf{D2~\cite{liu2023asvspoof}} & \textbf{D3~\cite{wang2024asvspoof}} & \textbf{D4~\cite{xie2025codecfake}} & \textbf{D5~\cite{frank2021wavefake}} \\
\midrule
\multirow{2}{*}{Pitch Shifting~\cite{chirkovskiy2025pitch}}  & 
MSE  & 0.0334 & 0.0218 & 0.005 & 0.0005 & 0.0038 &
\multirow{2}{*}{FGSM~\cite{goodfellow2014explaining}}  & 
MSE  & 0.0095 & 0.0098 & 0.0091 & 0.0046 & 0.0017 \\
 & SSIM & 0.887 & 0.872 & 0.954 & 0.947 & 0.919 & & SSIM & 0.946 & 0.95 & 0.969 & 0.957 & 0.952 \\
 \midrule
\multirow{2}{*}{Filtering~\cite{wu2020defense}} 
 & MSE  & 0.0235 & 0.0146 & 0.0032 & 0.0004 & 0.0025 &
\multirow{2}{*}{PGD~\cite{madry2017towards}} & 
MSE  & 0.0095 & 0.0098 & 0.0091 & 0.0046 & 0.0017 \\
 & SSIM & 0.819 & 0.776 & 0.927 & 0.911 & 0.857 & & SSIM & 0.947 & 0.952 & 0.969 & 0.958 & 0.953 \\
  \midrule
\multirow{2}{*}{Noise Addition~\cite{li2025measuring}}     
 & MSE  & 0.0002 & 0.0006 & 0.0001 & 0.0002 & 0.0001 &
\multirow{2}{*}{C \& W~\cite{carlini2017towards}} & 
MSE  & 0.0094 & 0.0098 & 0.009 & 0.0042 & 0.0017 \\
 & SSIM & 0.991 & 0.989 & 0.994 & 0.988 & 0.990 & & SSIM & 0.947 & 0.95 & 0.97 & 0.96 & 0.952 \\
  \midrule
\multirow{2}{*}{Quantization~\cite{grama2015quantization}}     
 & MSE  & 0.001 & 0.0011 & 0.0006 & 0.0003 & 0.0008 &
\multirow{2}{*}{DeepFool~\cite{moosavi2016deepfool}} & 
MSE  & 0.0069 & 0.0069 & 0.0072 & 0.0022 & 0.0012 \\
 & SSIM & 0.999 & 0.994 & 0.996 & 0.922 & 0.968 & & SSIM & 0.96 & 0.962 & 0.976 & 0.971 & 0.968 \\
\midrule
\multicolumn{2}{c}{\textbf{Avg MSE}}    & \textbf{0.0145} & \textbf{0.0095} & \textbf{0.0022} & \textbf{0.0004} & \textbf{0.0018} &
\multicolumn{2}{c}{\textbf{Avg MSE}}  & \textbf{0.0088} & \textbf{0.0091} & \textbf{0.0086} & \textbf{0.0039} & \textbf{0.0016} \\
\multicolumn{2}{c}{\textbf{Avg SSIM}}     & \textbf{0.924}  & \textbf{0.908}  & \textbf{0.968}  & \textbf{0.942}  & \textbf{0.933} &
\multicolumn{2}{c}{\textbf{Avg SSIM}} & \textbf{0.950}  & \textbf{0.954}  & \textbf{0.971}  & \textbf{0.962}  & \textbf{0.956} \\
\bottomrule
\end{tabular}
\end{adjustbox}
\label{tab:quality}
\vspace{-15pt}
\end{table}

\subsection{Results against Optimization-based AF Attacks} \vspace{-8pt}
We evaluated ADD robustness against four optimization-based AFs: FGSM, PGD, C\&W, and DeepFool (Section~\ref{sec:op_af}). Table~\ref{tab:opt_af} shows that FGSM~\cite{goodfellow2014explaining} dropped AUC to 0.35 and raised EER to 0.62, while PGD~\cite{madry2017towards} further degraded performance (AUC: 0.09, EER: 0.87). C\&W~\cite{carlini2017towards} and DeepFool~\cite{moosavi2016deepfool} also caused notable declines (avg. AUCs: 0.51 and 0.55). $R2$, $R3$, and $R6$ were highly vulnerable, whereas $R1$ and $R5$ showed better robustness.\\
Spectrogram-based ADDs were similarly impacted by optimization-based AFs (Table~\ref{tab:opt_af}). FGSM and PGD dropped AUC to ~0.35 and raised EER above 0.60. C \& W and DeepFool further degraded performance. While $S1$ and $S4$ showed slight robustness due to attention, $S2$ and $S5$ were moderately affected. These results reveal the need for robust models.
\vspace{-15pt}
\subsection{Defense against AF Attacks} \vspace{-8pt}
To counter the AFs described in Section~\ref{sec:af_methods}, we selected four top-performing ADDs: two raw-based ($R1$, $R5$) and two spectrogram-based ($S3$, $S4$) methods. These were trained with adversarial examples generated using random parameter selections. Table~\ref{tab:defense} shows the detection AUC and EER for both seen and unseen datasets. Compared to the results in Tables~\ref{tab:af_st} and \ref{tab:opt_af}, adversarial training improves the average AUC and EER to 0.76 and 0.28 for raw ADDs, and 0.83 and 0.18 for spectrogram-based ADDs. However, these values remain lower than the baseline ADD results, indicating the need for more robust ADDs to defend against AFs.\\
\vspace{-28pt}
\subsection{Qualitative Analysis}\vspace{-8pt}
To assess audio quality, we computed average mean square error (MSE) and structural similarity index measure (SSIM) between original and attacked audios (Table~\ref{tab:quality}). For statistical AFs, we used: pitch shift = –1 semitone, median filter size = 3, noise std = 0.001, and bit depth = 4. For optimization-based AFs, parameters were: FGSM $\epsilon$ = 0.001, PGD $\epsilon$ = 0.003, C \& W confidence = 10, DeepFool overshoot = 0.005, with $R4$ as the detection model.\\
As shown in Table~\ref{tab:quality}, noise and quantization cause less distortion than pitch shifting and filtering, which alter the audio's structure more noticeably. Still, MSE stays below 0.01 and SSIM above 0.90, indicating minimal overall distortion.\\
In contrast, optimization-based AFs introduce significantly less distortion compared to statistical AFs. The average MSE remains below 0.009 and SSIM above 0.95, which indicates high similarity to the input samples. Compared to statistical AF attacks, optimization-based methods generate AF attacks that preserve the audio quality better.
\vspace{-18pt}

\begin{figure}[!b]
    \centering
    \vspace{-15pt}
    \begin{tabular}{cccc}
    \includegraphics[width=2.6cm]{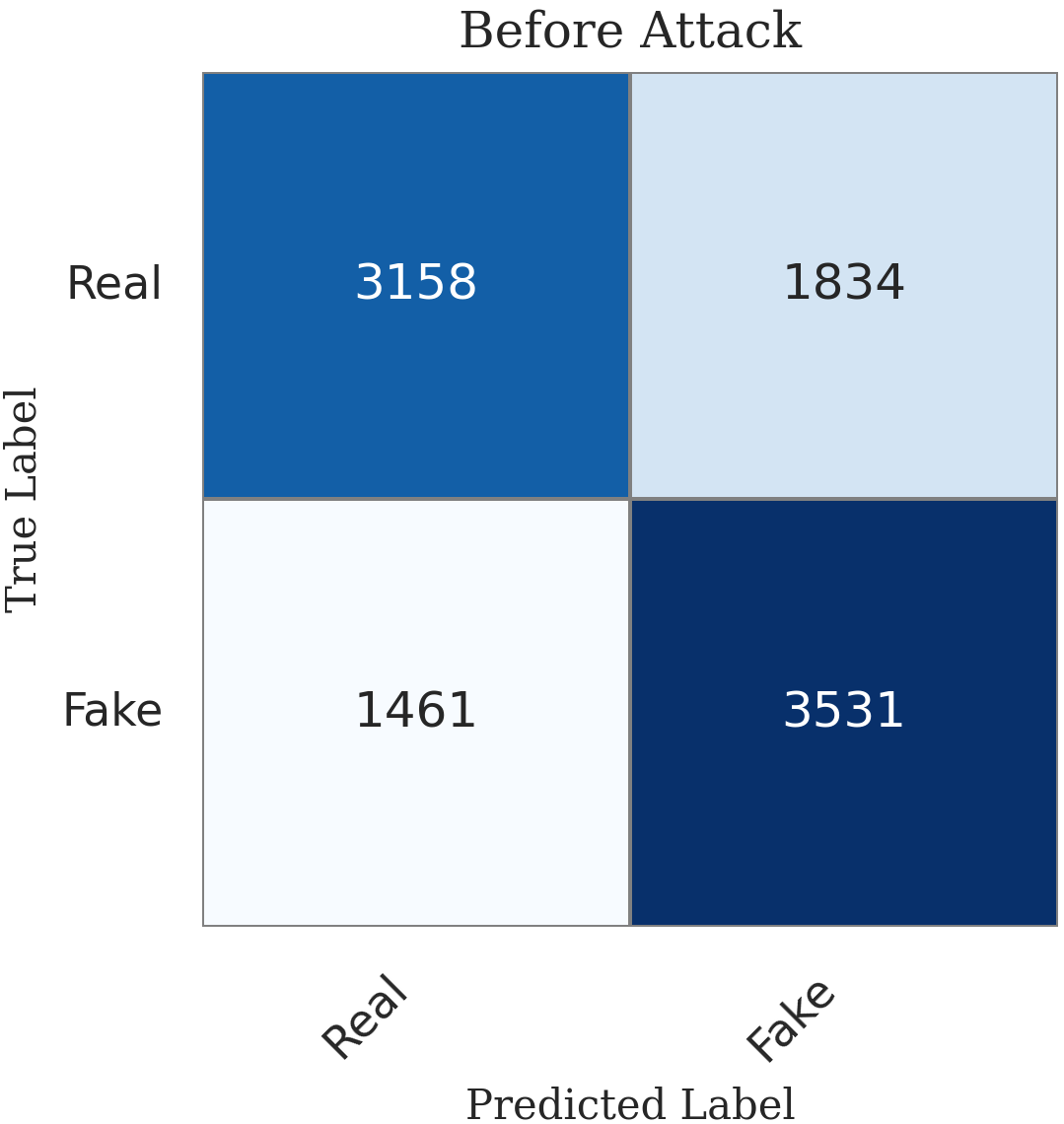} &
    \includegraphics[width=2.6cm]{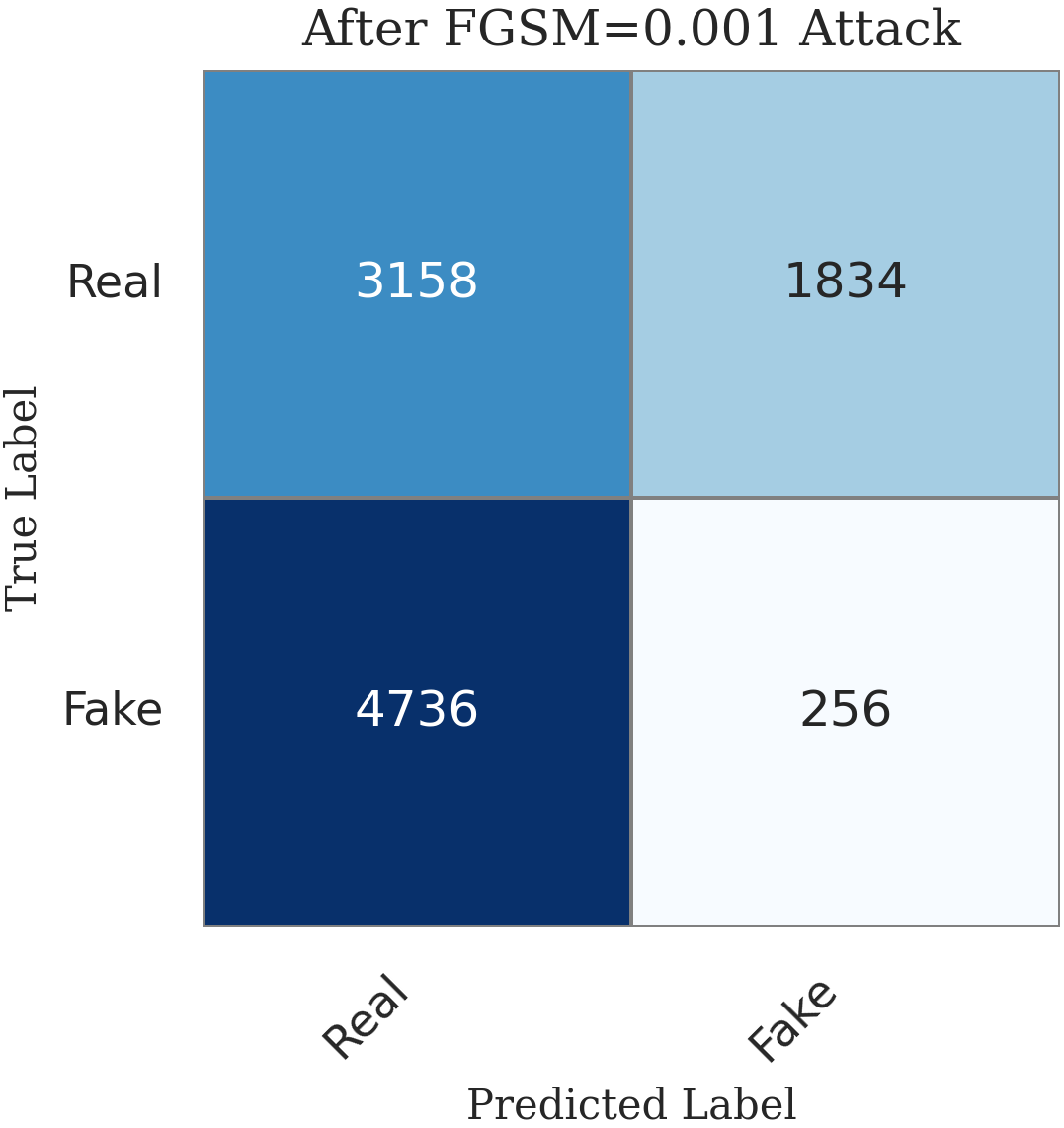} &
    \includegraphics[width=2.6cm]{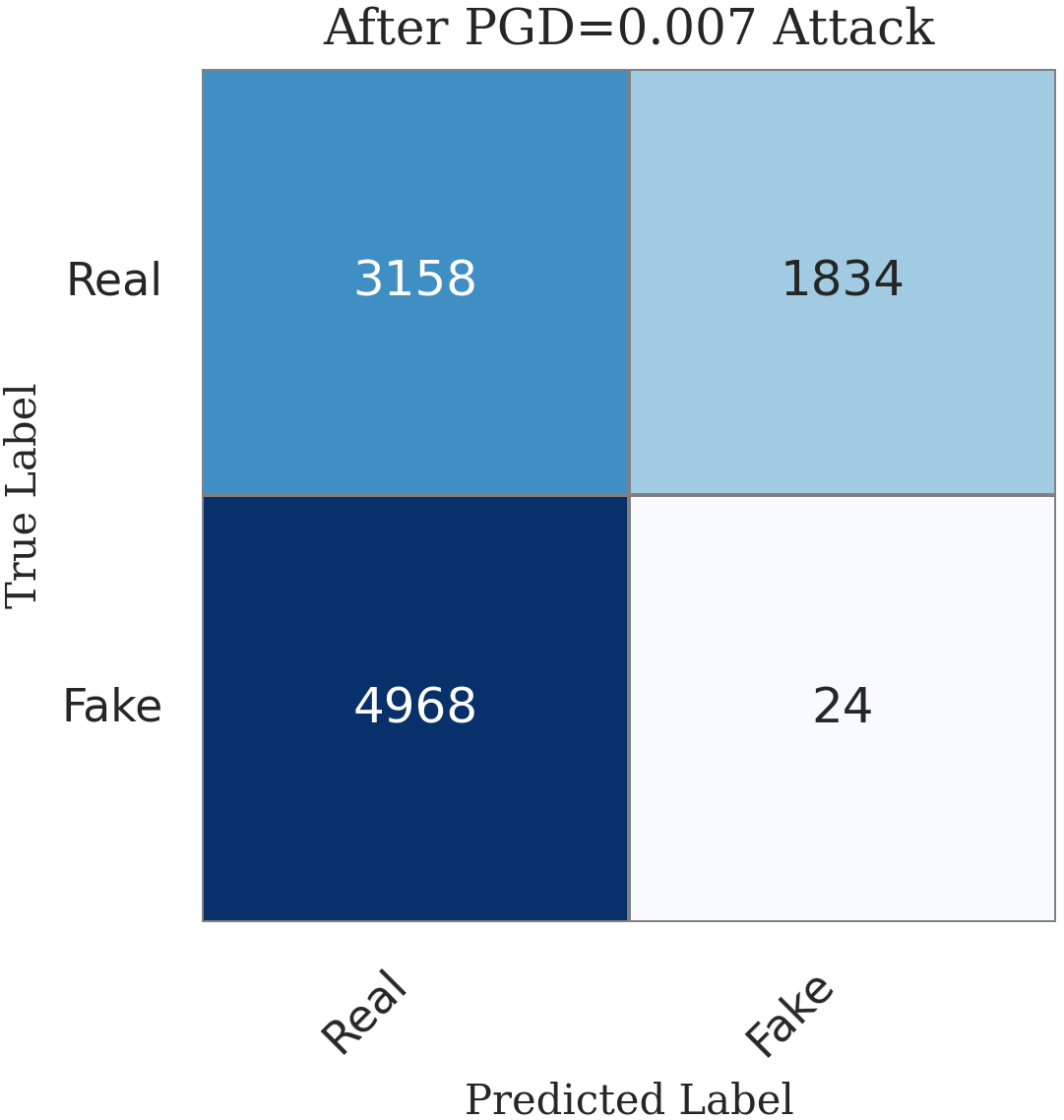} &
    \includegraphics[width=2.6cm]{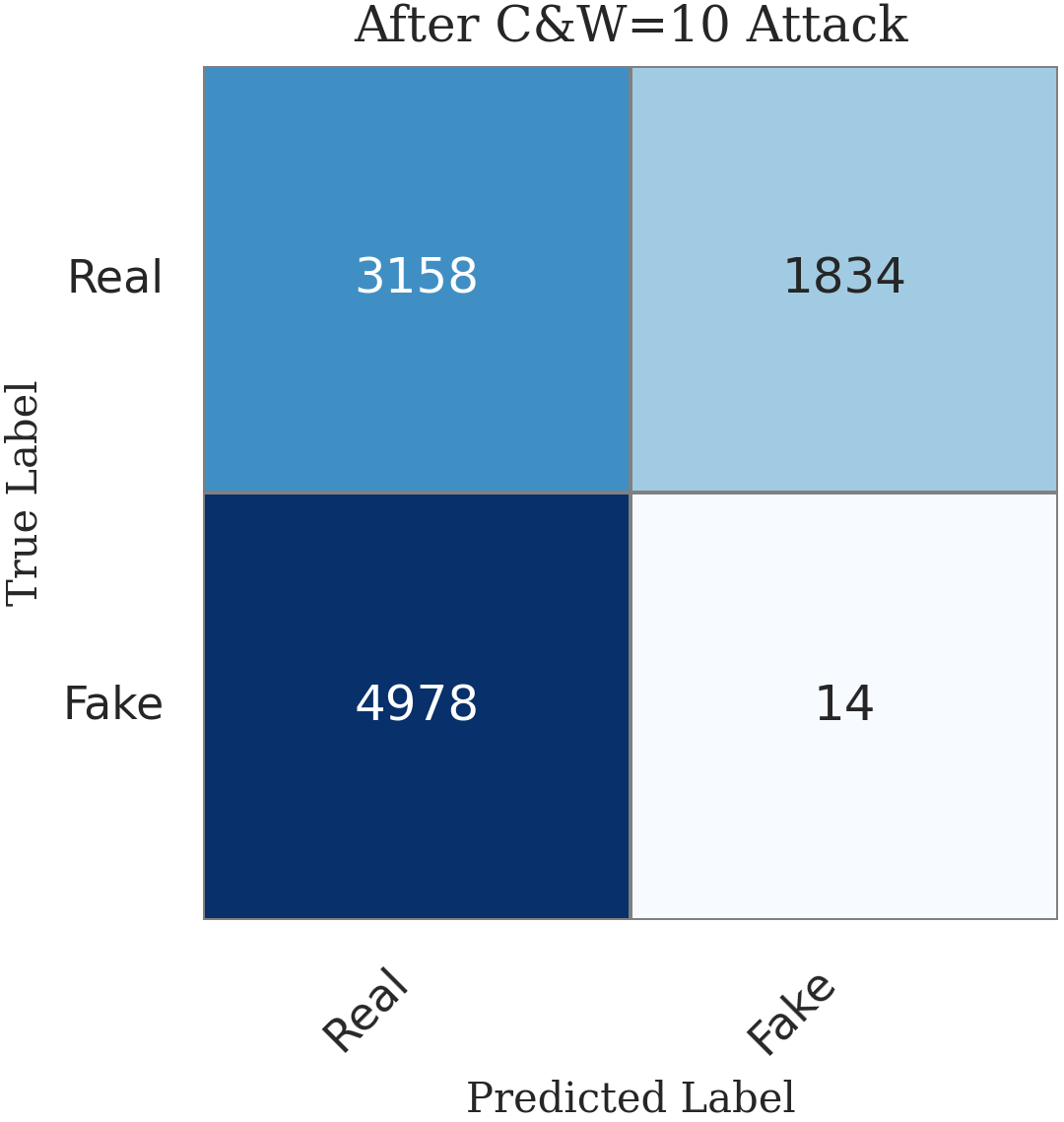} \\
    (a) & (b) & (c) & (d)
    \end{tabular}
    \vspace{2.5pt}
    \caption{Confusion matrix before and after AF attacks; (a) Original, (b) FGSM, (c) PGD, and (d) C \& W using $R4$ on the $D4$ dataset.}
    \label{fig:conf_mat}
    \vspace{-12pt}
\end{figure}

\begin{figure*}[!b]
    \centering
    % \begin{tabular}{cccc}
    \includegraphics[width=13cm, height=2.9cm]{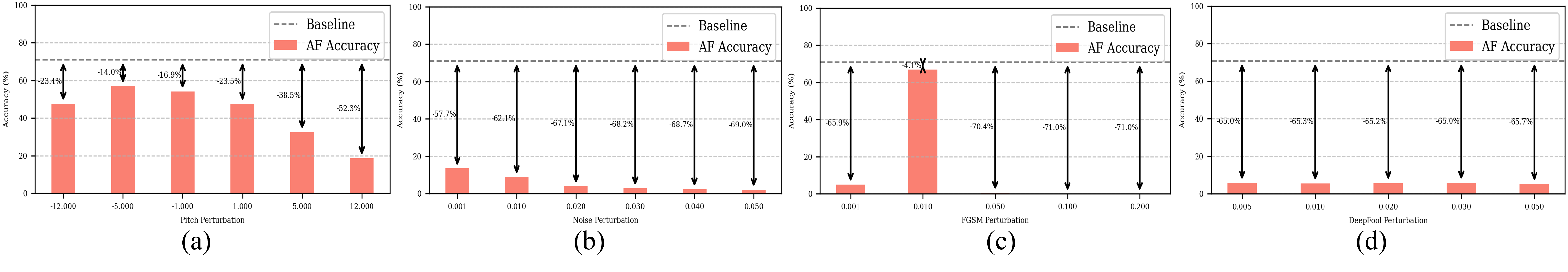} 
    % \includegraphics[width=2.93cm]{figures/noise.png} &
    % \includegraphics[width=2.93cm]{figures/fgsm.png} &
    % \includegraphics[width=2.93cm]{figures/df.png} \\
    % (a) & (b) & (c) & (d)
    % \end{tabular}
    % \vspace{0.1pt}
    \caption{Accuracy of AFs on ADD methods; (a) Pitch Shifting,  (b) Noise Addition, (c) FGSM, and (d) DeepFool with $R4$ on the $D4$ dataset.}
    \label{fig:st_eer}
    \vspace{-9pt}
\end{figure*}

\section{Discussions and Future Directions} \vspace{-8pt}
\subsection{Analysis and Discussions} 
\begin{figure}[!t]
    \centering
    \vspace{-2pt}
    \begin{tabular}{cccc}
    \includegraphics[width=6.2cm]{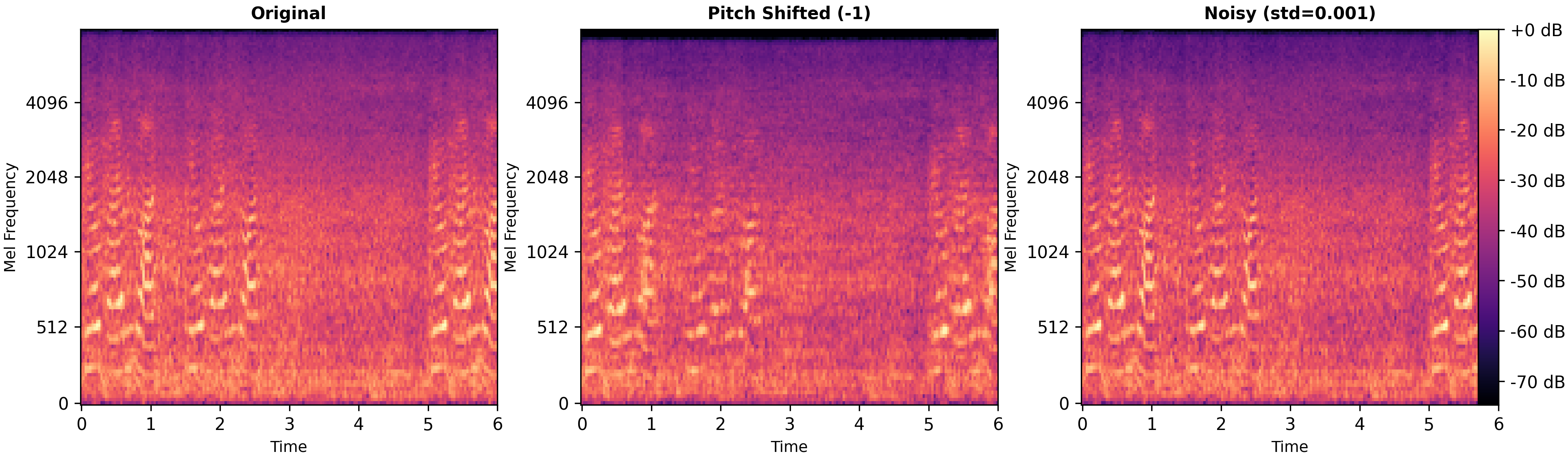} &
    \includegraphics[width=6.2cm]{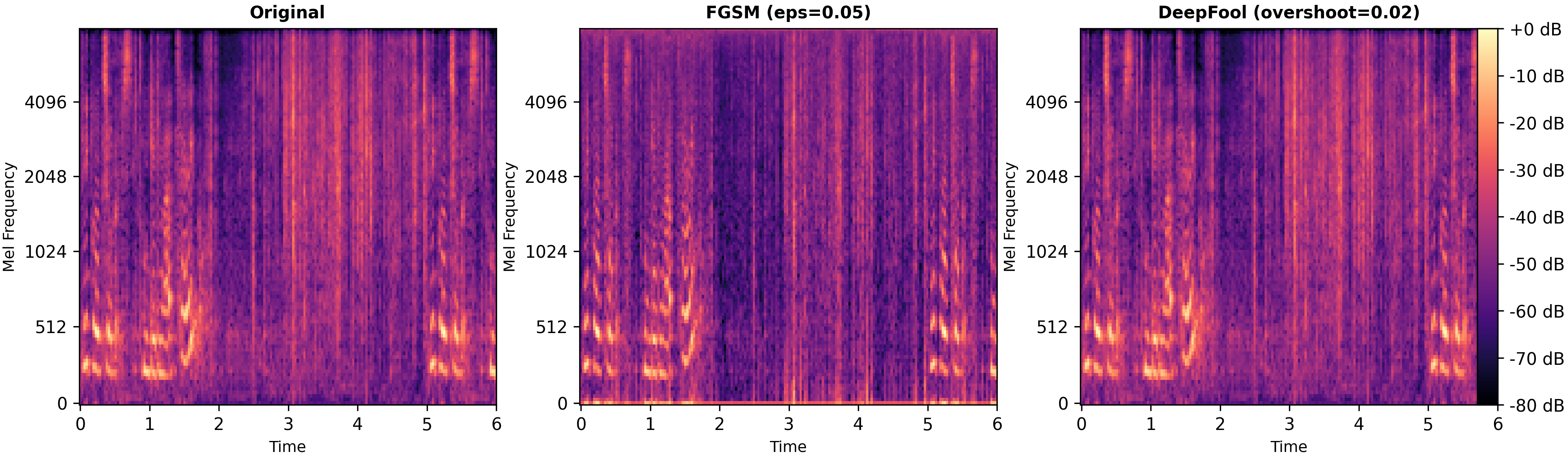}\\
    (a) & (b)
    \end{tabular}
    \vspace{2pt}
    \caption{Spectrogram visualization before and after AF attacks; (a) Original, Pitch, and  Noise, (b) Original, FGSM, and DeepFool.}
    \label{fig:st_spec}
    \vspace{-18pt}
\end{figure}

\vspace{-8pt}
This study provides a comprehensive evaluation of SoTA ADDs under various AF scenarios. The proposed comparative analysis highlights that while current raw and spectrogram-based ADDs perform reasonably well on clean data (randomly selected 10000 samples from $D5$), as shown in Figure~\ref{fig:conf_mat} (a), their effectiveness significantly degrades under AFs as depicted in Figure~\ref{fig:conf_mat} (b)-(d). Statistical AFs, such as pitch shifting, filtering, noise addition, and quantization, can obscure generative AI signatures as visualized in Figure~\ref{fig:st_eer} (a) \& (b). In contrast, optimization-based AFs (e.g., FGSM, DeepFool) expose fundamental vulnerabilities of ADDs by exploiting their targeted perturbations, as shown in Figure~\ref{fig:st_eer} (c) \& (d).\\
The observed performance drop, as shown in Figures~\ref{fig:st_eer} (a)-(d), underscores the critical challenge in designing generalized and resilient detectors that can adapt to diverse and evolving AF strategies. Furthermore, the qualitative analysis, including spectrogram visualization as given in Figure~\ref{fig:st_spec} (a) \& (b), reveals that some AF attacks introduce subtle distortions~\ref{fig:st_spec} (a) for pitch shifting that may go unnoticed by human listeners but are sufficient to deceive ADD models. These suggest that while spectrogram-based methods benefit from well-established backbone models, raw audio-based models retain advantages in capturing temporal dynamics, indicating that a hybrid approach could be beneficial.\\
Overall, these findings underscore the need to move beyond conventional supervised learning toward adaptive and explainable models that can effectively mitigate sophisticated AFs. \\

\vspace{-25pt}
\subsection{Challenges and Future Directions} \vspace{-8pt}
\textbf{Evolving Nature of AF Attacks:} This comparative analysis offers valuable insights into how different AF attacks affect ADD methods and helps researchers better understand AF behaviors and identify blind spots of detectors.\\
\textbf{Model Vulnerability:} A key challenge revealed by our study is that current ADDs exhibit architecture-specific vulnerabilities. This calls for future research into designing more resilient architectures and learning strategies that can adapt to diverse AF techniques.\\
\textbf{Standardized Evaluation Framework:} This study introduces a unified and reproducible evaluation pipeline by leveraging a wide range of ADD and AF techniques, providing a consistent foundation for future research and comparison within the ADD community.\\
\textbf{Foundation for Robust Model Design:} The performance drops under AF attacks guides researchers toward designing advanced techniques like adversarial training, data augmentation, and domain adaptation to enable more robust and generalized systems.\\
\textbf{Multimodal Expansion:} The findings encourage future studies to incorporate complementary multimodal (e.g., video, lip-sync) to enhance robustness in real-world applications.
\vspace{-18pt}
\section{Conclusions}\vspace{-8pt}
Nowadays, deepfake audio is posing substantial risks to voice biometrics and related applications. While several SoTA ADDs, including raw and spectrogram-based approaches, demonstrate promising performance in identifying generative AI artifacts, their effectiveness is critically challenged by diverse AFs. These attacks, particularly statistical (pitch shifting, filtering, noise addition, and quantization) and optimization-based methods (FGSM, PGD, C \& W, and DeepFool), effectively alter or conceal AI signatures. The proposed comparative analysis across benchmark datasets highlights the vulnerabilities of both raw and spectrogram-based ADDs under AFs. These findings emphasize the urgent need for developing more resilient and generalized ADDs capable of countering diverse AF techniques.\\
In future research, we aim to design adaptive and explainable comparative study that cover a wider range of forensic and AF models, including foundation model-based ADD methods and generative AF attacks. Exploring multi-modal approaches that combine audio and visual modalities will further enhance research directions. Additionally, integrating adversarial training and meta-learning techniques holds promise for improving generalization and robustness against unseen AF attack types.
\bibliography{egbib}

\end{document}